# Two-photon laser spectroscopy of antiprotonic helium and the antiproton-to-electron mass ratio

Masaki Hori[1,2], Anna Sótér[1], Daniel Barna[2,4], Andreas Dax[2], Ryugo Hayano[2], Susanne Friedreich[3], Bertalan Juhász[3], Thomas Pask[3], Eberhard Widmann[3], Dezső Horváth[4,5], Luca Venturelli[6], Nicola Zurlo[6]

[1]*Max-Planck-Institut für Quantenoptik, Hans-Kopfermann-Strasse 1, 85748 Garching, Germany*

[2]*Department of Physics, University of Tokyo, Hongo, Bunkyo-ku, Tokyo 113-0033, Japan*

[3]*Stefan Meyer Institut für Subatomare Physik, Boltzmanngasse 3, Vienna 1090, Austria*

[4]*KFKI Research Institute for Particle and Nuclear Physics, H-1525 Budapest, Hungary*

[5]*Institute of Nuclear Research (ATOMKI), H-4001 Debrecen, Hungary*

[6]*Dipartimento di Chimica e Fisica per l'Ingegneria e per i Materiali, Università di Brescia & Istituto Nazionale di Fisica Nucleare, Gruppo Collegato di Brescia, I-25133 Brescia, Italy*

**Physical laws are believed to be invariant under the combined transformations of charge, parity, and time reversal[1]. This *CPT symmetry* implies that antimatter particles have exactly the same mass and absolute value of the charge as their particle counterparts. Metastable antiprotonic helium ($\bar{p}\,\text{He}^+$) is a three-body atom[2] consisting of a normal helium nucleus, an electron in the ground state, and an antiproton occupying a Rydberg state with high principal and angular momentum quantum numbers $n\sim l+1\sim 38$. Here we report two-photon spectroscopy of antiprotonic helium, in which $\bar{p}\,^3\text{He}^+$ and $\bar{p}\,^4\text{He}^+$ isotopes are irradiated by two counter-propagating laser beams. This excites non-linear, two-photon transitions of the antiproton of the type $(n,l)\rightarrow(n-2,l-2)$ at the deep UV wavelengths $\lambda=139.8$, 193.0, and 197.0 nm, which partially cancels the Doppler broadening of the laser resonance caused by the thermal motion of the atoms. The resulting narrow spectral lines allowed us to measure three transition frequencies with fractional precisions of 2.3−5 parts in $10^9$. By comparing the results with three-body quantum electrodynamics calculations, we derived the antiproton-to-electron mass ratio as 1836.1526736(23), where the parenthetical error represents one standard deviation. This agrees with the proton-to-electron value known to a similar precision.**

Antiprotonic atoms (denoted $\bar{p}\,\text{X}^+$) can be readily synthesized in a given element X by replacing the atomic electrons with a negatively-charged antiproton[2]. The substitution takes place spontaneously when antiprotons are brought to rest in the substance concerned. However, these exotic atoms are usually destroyed within picoseconds by electromagnetic cascade mechanisms that result in the annihilation of the antiprotons in the nucleus of X. The $\bar{p}\text{He}^+$ atom alone retains microsecond-scale lifetimes even in dense helium targets. The extreme longevity is due to the fact that the antiprotons trapped in the $n\sim l+1\sim 38$ Rydberg states have almost no overlap with the nucleus, and furthermore cannot easily deexcite by Auger emission of the electron owing to its large ($I\sim 25$ eV) binding energy and the large multiplicity ($\Delta l$) of the necessary transition. The electron protects the antiproton during collisions with other helium atoms, making $\bar{p}\text{He}^+$ amenable to laser spectroscopy.

The energy levels of $\bar{p}\text{He}^+$ have been calculated by three-body QED calculations (ref. 3 and V.I. Korobov, personal communication) to precisions of $1\times 10^{-9}$. The calculated values now include relativistic and radiative recoil corrections up to order $m\alpha^6$ and nuclear size effects. The fractional measurement precision of single-photon laser spectroscopy experiments[4,5] of $\bar{p}\text{He}^+$, however, has always been limited to $10^{-7}$–$10^{-8}$ due to the Doppler broadening effect. As in normal atoms, the thermal motion of $\bar{p}\text{He}^+$ at temperature $T$ strongly broadens the measured widths of the laser resonances by $\approx \nu\sqrt{8kT\log 2/Mc^2}$, wherein $\nu$ denotes the transition frequency, $k$ the Boltzmann constant, $M$ the atom's mass, and $c$ the speed of light.

One way to reach precisions beyond this Doppler limit is provided by two-photon spectroscopy. For example, the 1$s$-2$s$ transition frequency in atomic H has been measured to a precision of $10^{-14}$ with two counterpropagating laser beams, each with a frequency corresponding to half the 1$s$-2$s$ energy interval. This arrangement cancels the Doppler broadening to first order[6]. A similar experiment has been proposed for antihydrogen ($\bar{\text{H}}$) atoms[7,8] which were recently confined in a magnetic trap[8]. It is normally difficult, however, to apply this to $\bar{p}\text{He}^+$ because of the small probabilities involved in the nonlinear transitions of the massive antiproton. Indeed, calculations[9] show that gigawatt-scale laser powers would be needed to excite them within the atom's lifetime against annihilation.

Nevertheless we here induced two-photon transitions of the antiproton of the type $(n,l)=(n-2,l-2)$ for the first time [Fig. 1(a)], by using the fact that this probability can be enhanced by factor $>10^5$ if the counterpropagating beams have frequencies $\nu_1$ and $\nu_2$, such that the virtual intermediate state of the two-photon transition lies within $\Delta\nu_d\sim 10$ GHz of a real state[10] ($n$-1,$l$-1). At resonance between the atom and the laser beam, the antiprotons are directly transferred between the parent and daughter states via the nonlinear transition, leaving the population in ($n$-1,$l$-1) unaffected. The first-order Doppler width is then reduced by a factor $|\nu_1-\nu_2|/(\nu_1+\nu_2)$.

In these experiments, transitions were selected for laser excitation between pairs of states with microsecond and nanosecond-scale lifetimes against Auger emission of the electron. At the two-photon resonance, Auger decay left a $\bar{p}\text{He}^{2+}$ ion behind. No longer protected by the electron in the way described above, the ion was rapidly destroyed in Stark collisions with other helium atoms. Charged pions emerged from the resulting antiproton annihilations, passed through an acrylic sheet and produced Cherenkov light which was detected by photomultipliers. The two-photon resonance condition between the counterpropagating laser beams and the atom was thus revealed as a sharp spike in the annihilation rate [Fig. 1(b)].

Even under these conditions of enhanced transition probability, MW-scale laser pulses of high spectral purity and low phase noise are needed to excite these two-photon transitions and avoid rapid dephasing of its amplitude[10]. For this [Fig. 1 (c)] we developed two sets of Ti:Sapphire lasers of pulse length 30-100 ns with amongst the smallest linewidths (~6 MHz) reported so far[11]. They were based on continuous-wave (cw) lasers of wavelengths 728-940 nm whose frequencies were measured to a precision of $<1\times10^{-10}$ using a femtosecond optical comb[12] locked to a global-positioning-satellite disciplined, quartz oscillator. This cw seed beam was pulse-amplified to the necessary 1 MW peak power using a Ti:Sapphire oscillator and amplifier. Spurious modulations in the pulsed laser frequency or "chirp" induced during this amplification are an important source of systematic error[5,13-15], and were measured using a heterodyne spectrometer[11]. The $<1.4\times10^{-9}$ precision of this laser system was verified[11] by measuring some two-photon transition frequencies in Rb and Cs at wavelengths of 778 and 822 nm.

It was essential to use helium targets of low enough density for the relaxations caused by collisions between $\bar{p}\mathrm{He}^+$ and other helium atoms that could inhibit the two-photon transition to remain small. This implied the use of antiprotons of low enough energy to be stopped in such targets within the volume irradiated by the 2-cm-diameter laser beams. The Antiproton Decelerator (AD) of CERN provided 200-ns-long pulsed beams of 5.3-MeV antiprotons [Fig. 1(c)]. Every 100 s, we decelerated some $7\times10^6$ antiprotons to ~70 keV by allowing them to pass through a 3-m-long radiofrequency quadrupole decelerator[4]. The beam was then transported by an achromatic magnetic beamline to the target chamber filled with $^4$He or $^3$He gas at temperature $T$~15 K and pressure $p$=0.8−3 mbar. At a time 2−8μs after the resulting $\bar{p}\mathrm{He}^+$ formation, horizontally-polarized laser beams of energy density ~1 mJ/cm$^2$ were simultaneously fired through the target in a perpendicular direction to the antiproton beam.

Fig. 1(b) shows the Cherenkov signal (solid blue line) as a function of time elapsed since the arrival of antiproton pulses at the target, averaged over 30 pulses which corresponds to ~$10^7$ $\bar{p}\mathrm{He}^+$ atoms. Lasers of wavelengths $c/v_1$=417 and $c/v_2$=372 nm were tuned to the two-photon transition (36,34)→(34,32) so that the virtual intermediate state lay $\Delta v_d$~6 GHz away from the real state (35,33). The above-mentioned annihilation spike corresponding to the two-photon transition can be seen at $t$=2.4 μs. When the 417-nm laser alone was tuned slightly (by 0.5 GHz) off the two-photon resonance condition (red line), the signal abruptly disappeared as expected. This indicates that the background from any Doppler-broadened, single-photon transitions is very small.

Fig. 2(b) shows the resonance profile measured by detuning the $v_1$ laser to $\Delta v_d$=-6 GHz, whereas $v_2$ was scanned between -1 and 1 GHz around the two-photon resonance defined by $v_1+v_2$ corresponding to a wavelength of 197.0 nm. The measured linewidth (~200 MHz) represents by far the highest spectral resolution achieved for an antiprotonic atom, and is more than an order of magnitude smaller than the Doppler- and power-broadened profile of the corresponding single-photon resonance (36,34)→(35,33) [Fig. 2 (a)] measured under the same target and laser power conditions. This allows us to determine the atomic transition frequency with a correspondingly higher precision. The remaining width is caused by the hyperfine structure; the 3-ns Auger lifetime of the daughter state (34,32); and power broadening effects.

The two-peak structure with a frequency interval of 500 MHz arises from the dominant interaction between the electron spin and the orbital angular momentum of the antiproton. Each peak is a superposition of two hyperfine lines caused by a further interaction between the antiproton and electron spins. The asymmetric structure is reproduced by lineshape calculations[9] (see below) and is due to the 48-MHz spacing between the unresolved hyperfine lines $(S_e, S_{\bar{p}})=(\uparrow\uparrow)\rightarrow(\uparrow\uparrow)$ and $(\uparrow\downarrow)\rightarrow(\uparrow\downarrow)$ being smaller than the 140-MHz one between $(\downarrow\uparrow)\rightarrow(\downarrow\uparrow)$ and $(\downarrow\downarrow)\rightarrow(\downarrow\downarrow)$.

We next detected the $(33,32)\rightarrow(31,30)$ resonance at wavelength $\lambda$=139.8 nm with the lowest $n$-values among the two-photon transitions, using lasers of $c/\nu_1$ =296 nm and $c/\nu_2$ =264 nm [Fig.2 (c)]. The small transition probability and antiproton population required that higher laser intensities $p$>2 mJ/cm$^2$ and small detunings $\Delta\nu_d$~3 GHz from state (32,31) were needed. All four hyperfine lines are much closer together (±100 MHz). We also measured the $\bar{p}^3$He$^+$ resonance $(35,33)\rightarrow(33,31)$ of $\lambda$=193.0 nm [Fig.2 (d)] using lasers of $c/\nu_1$ =410 nm and $c/\nu_2$ =364 nm. This profile contains eight partially-overlapping hyperfine lines arising from the spin-spin interactions of the $^3$He nucleus, electron, and antiproton.

We determined the spin-independent transition frequencies $\nu_{exp}$ (Table 1), by fitting each profile with a theoretical lineshape[9] (Fig. 2, blue lines) which was determined by numerically solving the non-linear rate equations of the two-photon process. This included all two-photon transitions between the $(2l+1)$ ~70 substates, the transition rates, power broadening effects, thermal motion of the atoms, the spurious frequency modulation[11] in the laser pulse, the experimentally-measured spatial and temporal profiles of the laser beam, and ac Stark effects[9]. The positions of the hyperfine lines were fixed to the theoretical values, which have a precision of <0.5 MHz (ref. 16).

For the transition $(36,34)\rightarrow(34,32)$ in $\bar{p}^4$He$^+$ (Table 2), the statistical error $\sigma_{stat}$ due to the finite number of atoms in the laser beam was estimated as 3 MHz (all errors quoted are standard deviations). Transitions were measured at various target densities between $1\times10^{18}$ and $3\times10^{18}$ cm$^{-3}$. Within this density range, no significant collisional shift was observable within the 3-MHz experimental error. This agrees with quantum chemistry calculations[17] for which the predictions of 0.1−1-MHz-scale collisional shifts in the associated single-photon lines agreed with experimental results[4,18] within ~20%. Calculations show that magnetic Zeeman shifts are also small <0.5 MHz for the Rydberg states under our experimental conditions. The frequency chirp of each laser pulse was recorded and corrected to a precision[11] of 0.8 MHz. The systematic error arising from the calculation of the fitting function was estimated[9] to be around 1 MHz.

Laser fields can shift the frequencies of the two-photon transitions[9] by an amount proportional to $(\Omega_1^2-\Omega_2^2)/\Delta\nu_d$, where $\Omega_1$ and $\Omega_2$ denote the Rabi frequencies of transitions between the parent and virtual intermediate states and, the daughter and intermediate states. We reduced this a.c. Stark shift to ≤5 MHz by carefully adjusting the intensities of the two laser beams such that $\Omega_1\sim\Omega_2$. Remaining shifts were canceled to a level of 0.5MHz by systematically comparing[9] the resonance profiles measured alternately at positive and negative detunings, $\pm\Delta\nu_d$. The total experimental error $\sigma_{exp}$ was obtained as the quadratic sum of all these errors. The larger error for the 193.0-nm $\bar{p}^3$He$^+$ transitions is due to the larger number (eight) of hyperfine lines and the smaller signal intensity.

The experimental transition frequencies $\nu_{\text{exp}}$ (Fig. 3, filled circles with error bars) agree with theoretical $\nu_{\text{th}}$ values (squares) within $(2-5)\times 10^{-9}$. This agreement is five to ten times better than previous single-photon experiments[5]. The calculation uses fundamental constants[19] compiled in CODATA2002 including the $^3$He-to-electron and $^4$He-to-electron mass ratios, the Bohr radius, and Rydberg constant. To preserve independence we avoided using the more recent CODATA 2006 values, which include results from our previous experiments and three-body QED calculations on $\bar{p}\text{He}^+$. The charge radii of the $^3$He and $^4$He nuclei give corrections to $\nu_{\text{th}}$ of 4−7 MHz, whereas the correction from the antiproton radius is much smaller[3,20] (<1 MHz) owing to the large $l$-value of the states. The precision of $\nu_{\text{th}}$ is mainly limited by the uncalculated radiative corrections of order $mc^2\alpha^8/h$ (Table 2).

When the antiproton-to-electron mass ratio $M_{\bar{p}}/m_e$ in these calculations was changed by $10^{-9}$, the $\nu_{\text{th}}$-value changed by 2.3-2.8 MHz. By minimizing $\sum \left[\nu_{\text{th}}(M_{\bar{p}}/m_e) - \nu_{\text{exp}}\right]_{\bar{p}}^2 / \sigma_{\text{stat}}^2$, where the sum is over the three $\bar{p}\text{He}^+$ frequencies, and considering the above systematic errors, $\sigma_{\text{sys}}$, we obtained the ratio $M_{\bar{p}}/m_e$=1836.1526736(23) which yielded the best agreement between theoretical and experimental frequencies. The uncertainty of $23\times 10^{-7}$ includes the statistical and systematic experimental, and theoretical contributions of $18\times 10^{-7}$, $12\times 10^{-7}$, and $10\times 10^{-7}$. This is in good agreement with the four previous measurements of the proton-to-electron mass ratio[21-24] (Fig. 4) with a similar experimental precision. The most precise value for protons is currently obtained by comparing the g-factors of hydrogen-like $^{12}\text{C}^{5+}$ and $^{16}\text{O}^{7+}$ ions measured by the GSI-Mainz collaboration[23,24] with high-field QED calculations. The CODATA recommended value for $M_p/m_e$ is taken as the average of these experiments. This ratio may be determined to a higher precision in the future by laser spectroscopy experiments[25] on $H_2^+$ and $HD^+$ ions. By assuming[19] CPT invariance $M_{\bar{p}}=M_p$=1.00727646677(10) u, we can further derive a value for the electron mass $m_e$=0.0005485799091(7) u from the $\bar{p}\text{He}^+$ result.

Hughes and Deutch[26,27] constrained the equality between the antiproton and proton charges and masses, formulated respectively as $\delta_Q=(Q_p+Q_{\bar{p}})/Q_p$ and $\delta_M=(M_p-M_{\bar{p}})/M_p$, to better than $2\times 10^{-5}$. For this they combined X-ray spectroscopic data of antiprotonic atoms ($\propto Q_{\bar{p}}^2 M_{\bar{p}}$) and the cyclotron frequency ($\propto Q_{\bar{p}}/M_{\bar{p}}$) of antiprotons confined in Penning traps measured to a higher precision. We can improve this limit by more than 4 orders of magnitude, by studying the linear dependence[2] of $\delta_M$ and $\delta_Q$ on $\nu_{\text{th}}$, i.e., $\delta_M\kappa_M+\delta_Q\kappa_Q<|\nu_{\text{exp}}-\nu_{\text{th}}|/\nu_{\text{exp}}$. For the three transitions, the constants were estimated[2] as $\kappa_M$=2.3-2.8 and $\kappa_Q$=2.7-3.4, whereas the right side of this equation was evaluated by averaging over the three transitions as $<(8\pm 15)\times 10^{-10}$. Meanwhile the constraint of $(Q_{\bar{p}}/M_{\bar{p}})/(Q_p/M_p)+1 = 1.6(9)\times 10^{-10}$ from the TRAP experiment[28,29] implies that $\delta_Q\sim\delta_M$. From this we conclude that any deviation between the charges and masses are $<7\times 10^{-10}$ at the 90% confidence level.

**Methods Summary.** The two cw seed lasers were stabilized relative to a 470-mm-long, monolithic cavities made of ultra-low expansion (ULE) glass by utilizing the Pound-Drever-Hall technique. The cavities were suspended horizontally by springs and isolated in a vacuum chamber whose temperature was stabilized to ± 0.05°C. Drifts in the laser frequencies were typically <0.1 MHz/h. The frequency chirp[11,13-15] during pulsed laser amplification was corrected using electro-optic modulators placed inside the pulsed laser resonators so that its amplitude was reduced to a few MHz. The remaining chirp was recorded for each laser pulse and its effect corrected for at the data analysis stage. The output beams were frequency doubled (SHG) or tripled (THG) to wavelengths $\lambda$=264−417 nm in beta-barium borate and lithium triborate crystals. Simulations[5,13,14] show that additional chirp caused by this frequency conversion is negligible (<0.1 MHz).

The Cherenkov signals corresponding to $\bar{p}\text{He}^+$ were recorded using a digital oscilloscope, and the area under the peak in each of these spectra (Fig. 1b) was plotted as a function of laser frequency to obtain the resonance profiles in Fig 2. Each data point represents an average of 8−10 antiproton beam arrivals at the target. This measurement was repeated over 10000 arrivals at various laser intensities, target densities, frequency offsets $\nu_d$, and alignments of the antiproton beam. The fact that the pulsed laser can maintain the absolute precision over a similar period of time was verified by using part of the light to measure the 6s-8s two-photon transition frequency of caesium 20 times over a 2-week period. The result with a conservative error of $1.4\times10^{-9}$ was in good agreement with previous experiments[30]. The acquired resonance profile was fitted with the theoretical two-photon resonance lineshape as described in the main text. This apriori calculation well reproduced the experimental data (Fig. 2). The validity of this method was also partially verified by using it to analyze the above-mentioned caesium two-photon signal[11].

**Table 1: Spin-averaged transition frequencies of $\bar{p}\text{He}^+$: experimental values with total, statistical, and systematic one standard deviation errors in parenthesis; theoretical values with uncertainties from uncalculated QED terms and numerical errors.**

| transition | Transition frequency (MHz) | |
|---|---|---|
| (n,l)→(n-2,l-2) | Expt. | Korobov[3] |
| $\bar{p}\,^4\text{He}^+$ | | |
| (36,34)→ (34,32) | 1522107062(4)(3)(2) | 1522107058.9(2.1)(0.3) |
| (33,32)→ (31,30) | 2145054858(5)(5)(2) | 2145054857.9(1.6)(0.3) |
| $\bar{p}\,^3\text{He}^+$ | | |
| (35,33)→ (33,31) | 1553643100(7)(7)(3) | 1553643100.7(2.2)(0.2) |

**Table 2: Experimental and theoretical one standard deviation errors associated with transition (*n,l*)=(36,34)→(34,32) of $\bar{p}\,^4\text{He}^+$.**

|  | Error (MHz) |
|---|---|
| Statistical error $\sigma_{\text{stat}}$ | 3 |
| Collisional shift error | 1 |
| AC Stark shift error | 0.5 |
| Zeeman shift | <0.5 |
| Frequency chirp error | 0.8 |
| Seed laser frequency calibration | <0.1 |
| Hyperfine structure | <0.5 |
| Line profile simulation | 1 |
| Total systematic error $\sigma_{\text{sys}}$ | 1.8 |
| Total experimental error $\sigma_{\text{exp}}$ | 3.5 |
|  |  |
| Error from uncalculated QED terms[3] | 2.1 |
| Numerical error of calculation[3] | 0.3 |
| Mass uncertainties[3] | <0.1 |
| Charge radii uncertainties[3] | <0.1 |
| Total theoretical error[3] $\sigma_{\text{th}}$ | 2.1 |

**Acknowledgements.** This work was supported by the European Science Foundation (EURYI), Monbukagakusho (grant no 20002003), the Munich Advanced Photonics cluster of Deutsche Forschungsgemeinschaft, Hungarian Research Foundation (K72172), and the Austrian Federal Ministry of Science and Research. We thank the CERN AD and PS operational staff, the CERN cryogenics laboratory, J. Alnis, D. Bakalov, J. Eades, R. Holzwarth, V.I. Korobov, M. Mitani, W. Pirkl, and T. Udem.

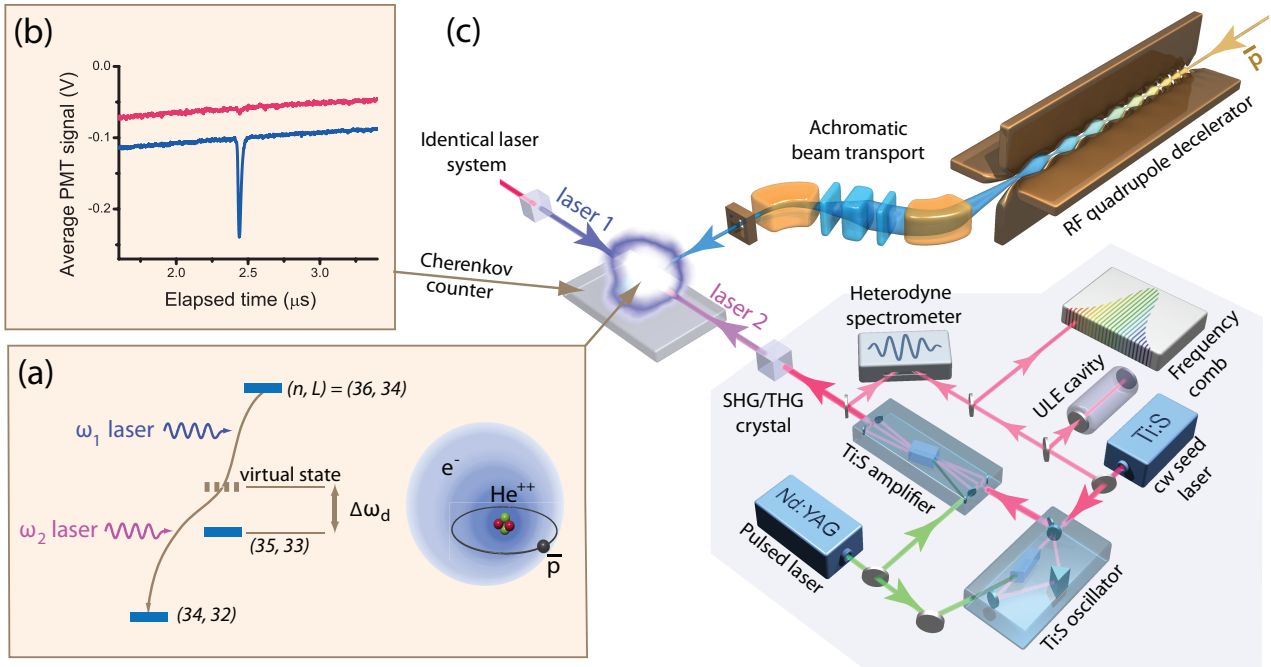

**Figure 1**: **Energy levels, Cherenkov detector signals, and experimental layout for two-photon spectroscopy of $\bar{p}\mathrm{He}^+$. a,** Two counter-propagating laser beams induced the two-photon transition $(n,l)=(36,34)\rightarrow(34,32)$ in $\bar{p}^4\mathrm{He}^+$ via a virtual intermediate state of the antiproton tuned close to the real state (35,33). **b**, Cherenkov detectors revealed the annihilation of $\bar{p}^4\mathrm{He}^+$ following the non-linear two-photon resonance induced at $t=2.4$ µs (blue). When one of the lasers was detuned -500 MHz away from resonance condition (red), the two-photon signal abruptly disappeared. PMT, photomultiplier tube. **c**, The $\bar{p}^4\mathrm{He}^+$ were synthesized by decelerating a beam of antiprotons using a radiofrequency quadrupole, and allowing them to stop in a cryogenic helium target. Two Ti:sapphire pulsed lasers whose optical frequencies were stabilized to a femtosecond comb were used to carry out the spectroscopy. CW, continuous wave; RF, radio frequency; SHG, second harmonic generation; THG, third-harmonic generation; ULE, ultralow expansion.

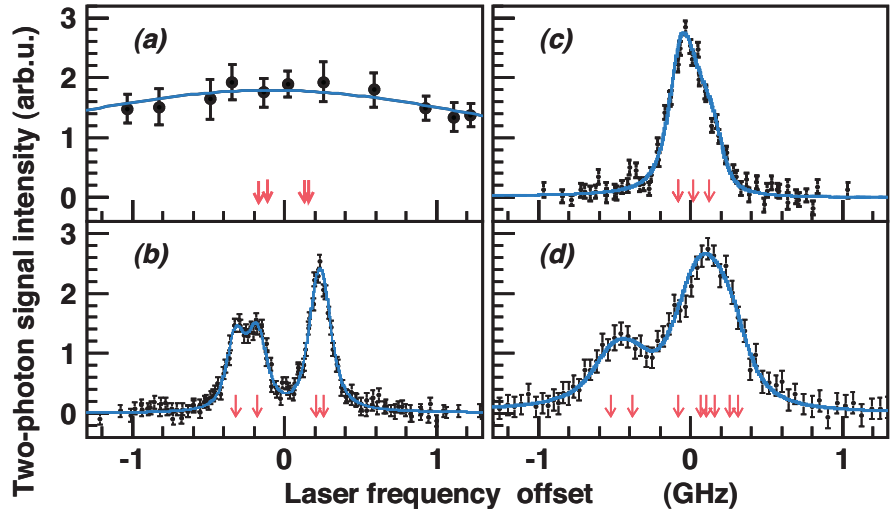

**Figure 2. Profiles of sub-Doppler two-photon resonances. a,** Doppler-and power-broadened profile of the single-photon resonance (36,34)→(35,33) of $\bar{p}^4\mathrm{He}^+$. **b,** Sub-Doppler two-photon profile of (36,34)→(34,32) involving the same parent state. **c,** Profiles of (33,32)→(31,30) of $\bar{p}^4\mathrm{He}^+$ and **d,** (35,33)→(33,31) of $\bar{p}^3\mathrm{He}^+$. Partially overlapping arrows indicate positions of the hyperfine lines. Error bars, 1 s.d; a.u., arbitrary units

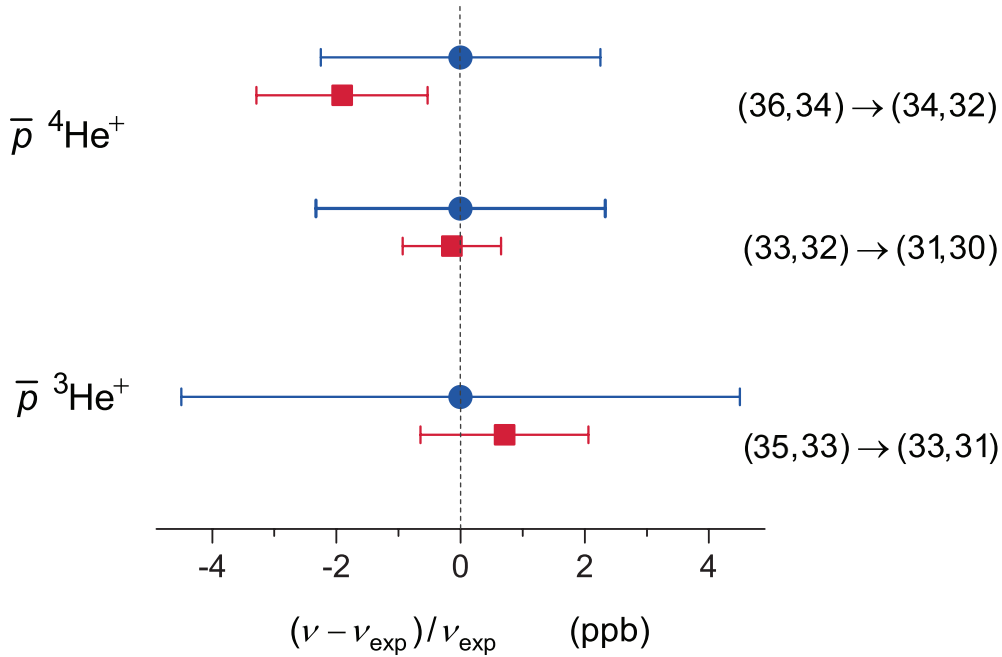

**Figure 3. Two-photon transition frequencies.** The experimental values (blue circles) for $\bar{p}^4\mathrm{He}^+$ and $\bar{p}^3\mathrm{He}^+$ agree with theoretical values (red squares) within fractional precisions of $(2-5)\times 10^{-9}$. Error bars, 1 s.d.; ppb., parts per billion.

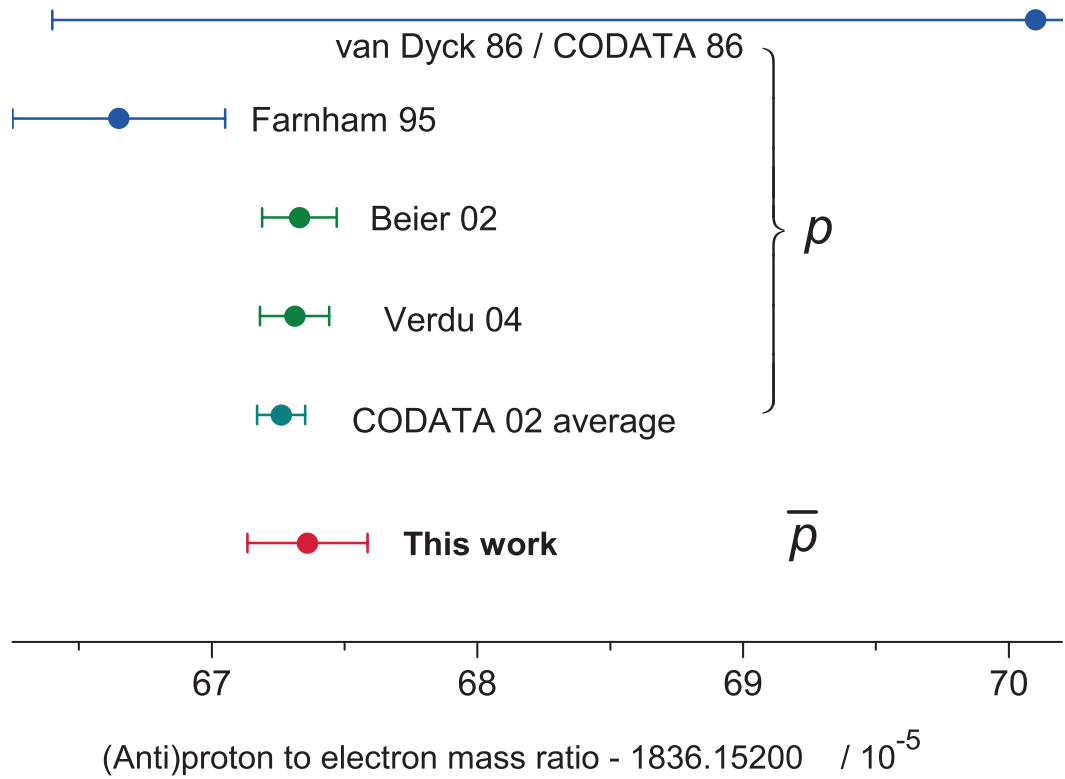

**Figure 4. Antiproton- and proton-to-electron mass ratios.** The antiproton-to-electron mass ratio determined by the present work agrees within a fractional precision of <1.3 part per billion with the proton-to-electron values measured in previous experiments[21-23] and the CODATA 2002 recommended value obtained by averaging them[20]. Error bars, 1 s.d.